# Twist Angle mapping in layered WS$_2$ by Polarization-Resolved Second Harmonic Generation

Sotiris Psilodimitrakopoulos[1], Leonidas Mouchliadis[1], Ioannis Paradisanos[1,2], George Kourmoulakis[1,3] Andreas Lemonis[1], , George Kioseoglou[1,3] and Emmanuel Stratakis[1,3]

[1]Institute of Electronic Structure and Laser, Foundation for Research and Technology-Hellas, Heraklion Crete 71110,Greece

[2]Department of Physics, University of Crete, Heraklion Crete 71003, Greece

[3]Department of Materials Science and Technology, University of Crete, Heraklion Crete 71003, Greece

Correspondence: E. Stratakis, E-mail: stratak@iesl.forth.gr

**Abstract**

Stacked atomically thin transition metal dichalcogenides (TMDs) exhibit fundamentally new physical properties compared to those of the individual layers. The twist angle between the layers plays a crucial role in tuning these properties. Having a tool that provides high-resolution, large area mapping of the twist angle, would be of great importance in the characterization of such 2D structures. Here we use polarization-resolved second harmonic generation (P-SHG) imaging microscopy to rapidly map the twist angle in large areas of overlapping WS$_2$ stacked layers. The robustness of our methodology lies in the combination of both intensity and polarization measurements of SHG in the overlapping region. This allows the accurate measurement and consequent pixel-by-pixel mapping of the twist angle in this area. For the specific case of 30$^o$ twist angle, P-SHG enables imaging of individual layers.

**Introduction**

The graphene-related atomically thin 2D TMDs show great promise for high-tech optoelectronic applications[1-4]. In particular, they exhibit unique nonlinear optical properties owing to their reduced dimensionality and lack of centrosymmetry, that give rise to pronounced SHG[5-9]. Analysis of the emitted SHG signal provides information on the crystal orientation[5-7] and homogeneity[8-9] as well as the thickness[10] and stacking sequence[11,12] of TMD structures. At the same time, there has been an increasing scientific interest in twisted TMD structures that can either occur during chemical vapour deposition (CVD) growth or be prepared artificially. In the latter case, one can tailor the interlayer coupling, that is strongly twist angle–dependent, and thus reveal new physical phenomena[13-16]. Therefore, the twist angle can be regarded as a new degree of freedom, enabling tuning of the physical properties of stacked 2D materials. A prominent example was recently reported in a ground-breaking work, showing that bilayer graphene exhibits unconventional superconductivity for a small value of the twist angle between the layers[17]. It has also been demonstrated that the twist angle allows control of the valley and band alignment of stacked 2D TMDs[11] and it results in ultraflat bands and shear solitons in twisted bilayer $MoS_2$[18], thus enabling ultrafast charge transfer between the 2D layers[19]. The strain induced SHG[20], the twist angle–dependent moiré-templated strain patterning[21], the interlayer valley excitons in TMD heterobilayers[22,23], the twist angle-dependent conductivities across $MoS_2$/graphene heterojunctions[24], and the moiré excitons in heterobilayers[25-28] are just a few more studies that have also been reported recently. These observations indicate the strong potential to harness and tune the physical properties of layered 2D materials, via the adjustment of the twist angle of the stacked layers[29]. In this context, the development of an optical technique capable to map the twist angle with high precision over large areas would be an invaluable tool for the construction and characterization of such new materials.

Earlier studies on the SHG interference from artificially stacked TMD bilayers have shown that the differences in SHG intensity can be attributed to differences in the armchair orientation between the two twisted TMD layers[12]. In addition, phase-resolved SHG techniques have also been used for the determination of the relative orientation between monolayers[30]. In all these cases however, solely variations in SHG intensity were used to identify the armchair angle difference, a criterion that cannot unambiguously exclude other phenomena as a source of these changes, such as structural transformations and inhomogeneities. The novelty of our method lies in the combination of intensity and polarization-resolved SHG measurements in the overlapping region of stacked 2D materials. SHG intensity-only measurements are insufficient for the determination of the twist angle since variations in intensity may also be caused by changes in the stacking sequence of the layers (e.g. from 2H to 3R). On the other hand P-SHG modulation may be due to imperfections of the crystal quality that result in local changes of the main crystallographic axis[9]. Our work aims to resolve this issue by offering, for the first time, a combined SHG intensity and P-SHG study of stacked 2D structures, that allows the pixel-by-pixel mapping of the twist angle.

As a proof of concept, we demonstrate the advantages of our all optical nonlinear imaging method, by application both to a pair of overlapping CVD-grown $WS_2$ layers with different armchair orientation, and an artificially dry stacked $WS_2$ twisted bilayer that has been produced with mechanical exfoliation. Interestingly, we show that there is a specific twist angle of 30°, for which the SHG signal originating from the stacked SLs can be selectively switched-on and -off. This enables the effect of optical discrimination of atomically thin layers and therefore provides a form of axial super-resolution SHG imaging of each individual layer of stacked 2D TMDs.

**Results and discussion**

Within our approach, the SHG emission from a layered TMD is considered as the result of interference between single-armchair layers (SLs). As a SL is considered a 2D layer of 2H or 3R stacking sequence[29, 31]. Each one of those SLs is acting as a surface phase array antenna inside the SHG active volume[32]. In order to examine the effect of the twist angle of the stacked SLs on their combined SHG pattern, we have performed polarization-in, polarization-out SHG measurements on a raster-scanned area of a CVD-grown $WS_2$ sample and an artificially prepared $WS_2$/ $WS_2$ bilayer.

In our recent work on P-SHG of TMDs, we have demonstrated that the SHG signal modulates as the angle of linear polarization of the excitation field, $\varphi$, rotates and that the modulation depends on the armchair orientation. Using this P-SHG modulation one can map with high precision (~0.15°) the armchair angle, $\theta$, at every pixel of the image[9]. In the case of two SLs that partially overlap, the P-SHG from each individual layer modulates with a phase depending on the armchair orientation of the layer, while in the overlapping area, the P-SHG modulation follows the SHG interference of the two SLs. Based on this effect, it can be shown that upon using appropriate linear polarization for the excitation of layered $WS_2$ with two different armchair angles overlapping in a region, we are able to decompose the interfered SHG signal originating from each individual layer. This is shown in Fig. 1 and Video S1, presenting the P-SHG modulation measured from a multilayered CVD-grown $WS_2$ triangular flake that exhibits a central region of enhanced SHG intensity. In particular, by rotating the linear polarization of the excitation field, incident to the overlapping area of two SLs, we are able to switch–on the SHG signal from one SL, while the SHG signal from the other is switched–off and vice versa. As shown in detail below, such SHG switch–on/off effect occurs when the armchair angle difference, i.e. the twist angle, between the layers equals to 30°.

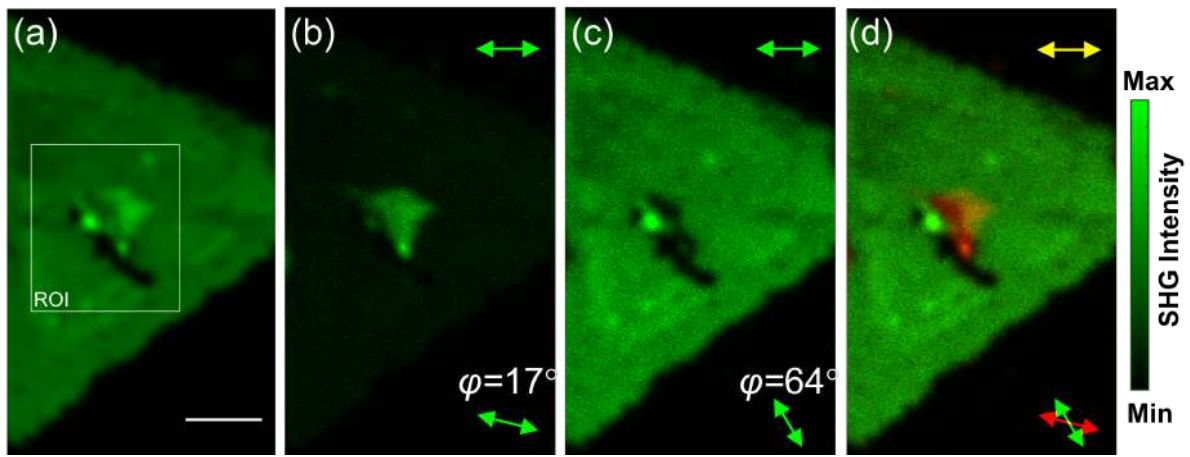

**Figure 1**: P-SHG–based identification of SLs with different armchair orientations. Selective SHG imaging of stacked SLs results in their optical isolation and provides sub-diffraction SL identification; (a) Total SHG intensity in the absence of an analyzer. The scale-bar is 5μm; (b) For $\varphi$= 17° we identify a SL. The double-headed arrows show the orientation of excitation linear polarization $\varphi$ (lower right corner) and the orientation of analyzer axis (upper right corner), respectively; (c) For $\varphi$= 64° we optically isolate a second SL; (d) Superimposed SHG intensities from the two different SLs (red and green for $\varphi$= 17° and $\varphi$= 64°, respectively).

At the stacking level, the produced SHG originates from the interference between at least two stacked SLs. It is therefore determined by the relative orientation of the two SLs, i.e., the twist angle. In a CVD-grown TMD, as in the case of our $WS_2$ sample, the crystal could be a mixture of 3R and 2H phases[33]. A direct consequence of such deviation from the ideal stacking sequence is the incomplete constructive or destructive interference of the SHG fields from different layers. Furthermore, the surface dipoles of the SLs are misaligned and the total SHG signal, as well as its polarization, depend on the twist angle. Regardless such crystal structure deviations, here we show that by using P-SHG measurements one can map for each pixel the armchair orientation of each SL constituting the multilayered structure. In order to accomplish that, the measurement utilizes the rotation of the linear excitation polarization with respect to the X-lab axis and a polarization analyzer parallel to X-lab axis, prior to SHG signal detection (Fig. 2). The armchair orientation of the first SL ($x_1y_1z_1$ coordinate system) is at an angle $\theta_1$ with respect to the X-axis, whereas the armchair orientation of the second SL ($x_2y_2z_2$ coordinate system) is at an angle $\theta_2$ with respect to the same axis (Fig. 2). In the polarization-in polarization-out measurements, the propagation of the laser beam is along $Z = z_1 = z_2$.

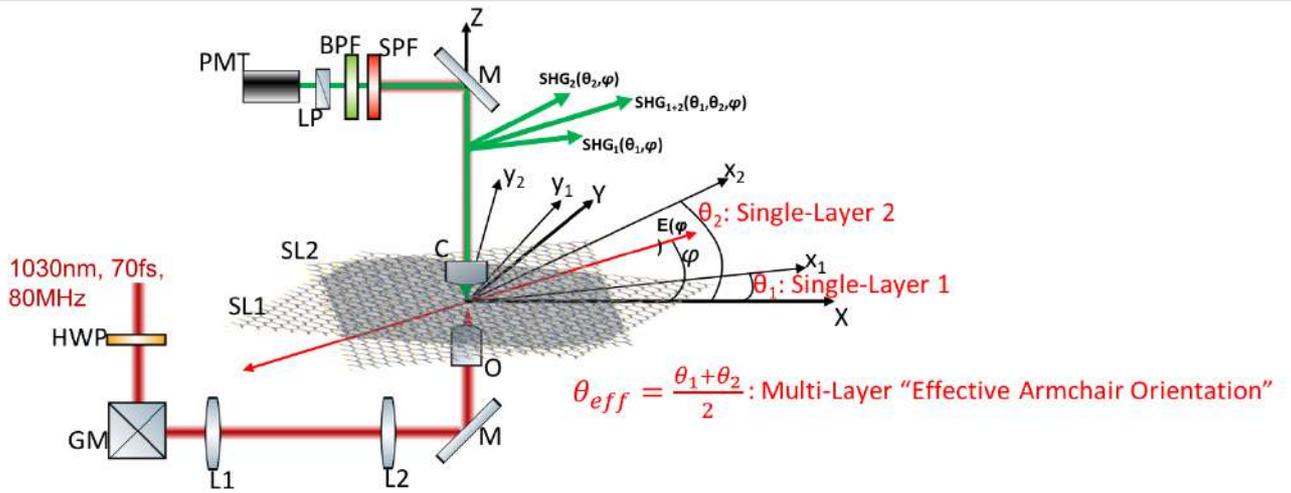

**Figure 2:** Coordinate systems and experimental setup for polarization-in, polarization-out measurements; HWP: half-waveplate; GM: galvanometric mirrors; L1, L2: lenses; M: mirror at 45°; O: objective; SL1, SL2: single-armchair layers; C: condenser; SPF: short-pass filter; BPF: bandpass filter; LP: linear polarizer; PMT: photomultiplier tube; XYZ: Lab coordinate system; $x_1y_1Z$ and $x_2y_2Z$: coordinate systems of SL1 and SL2, respectively; $\theta_1$, $\theta_2$: angles between XYZ and $x_1y_1Z$, $x_2y_2Z$, respectively; $\varphi$: angle between X-axis and excitation linear polarization $E(\varphi)$. We note three distinct

regions: clear SL1, clear SL2 and their overlapping region producing $SHG_1$ $(\theta_1,\varphi)$, $SHG_2(\theta_2,\varphi)$ and $SHG_{1+2}(\theta_1, \theta_2,\varphi)$, respectively.

Based on our previous work[9], it is straightforward to obtain an expression for the detected SHG signal in the case of two overlapping SLs (Fig. 3). The total SHG field in the overlapping area is given by the vector sum of the SHG signal from each layer. For the case of the analyzer orientation parallel to the lab X-axis and rotating linear polarization $\varphi$ the recorded SHG intensity is described by:

$$I_s = |A_1 \cos(3\theta_1 - 2\varphi) + A_2 \cos(3\theta_2 - 2\varphi)|^2 \qquad (1)$$

Here, $\theta_1$, $\theta_2$ correspond to the armchair orientations of the SL1 and SL2 respectively, and $A_i = C_i \, \varepsilon_0 \, \chi^{(2)}_{xxx}$, with $C_i$ being constants that depend on the local fields and the number of monolayers comprising the SL[33].

We note from Eq. (1) that for $A_1 = A_2 = A$, and for $\begin{cases} 3\theta_1 - 2\varphi = 0° \\ 3\theta_2 - 2\varphi = 90° \end{cases}$ or $\begin{cases} 3\theta_1 - 2\varphi = 90° \\ 3\theta_2 - 2\varphi = 0° \end{cases}$, in the overlapping region of two monolayers we obtain SHG equal to the SHG from each individual monolayer. This results in a form of axial super-resolution imaging of individual 2D layer and occurs at the 'magic'-SHG twist angle: $\delta = \theta_1 - \theta_2 = \pm 30°$. The total detected SHG intensity from both SLs, is also given by:

$$I_s = 4 A^2 \left(\cos\tfrac{3}{2}\delta\right)^2 \left(\cos(3\theta_{eff} - 2\varphi)\right)^2, \qquad (2)$$

where $\theta_{eff} = \frac{\theta_1+\theta_2}{2}$ is the effective armchair orientation in the overlapping region. Since $\theta_1, \theta_2 \in [0°,60°]$ we have that $\delta \in [-60°,60°]$ and $\theta_{eff} \in [0°,60°]$.

Note that the SHG intensity depends on the armchair angle difference δ between the SLs, being maximum for δ= 0° and zero for δ= 60°, while for δ≤ 30° and δ≥ 30° we have partially constructive and partially destructive SHG interference, respectively. The SHG of $N$ number of such SLs for the case of the analyzer orientation parallel to X-axis and rotating linear polarization $\varphi$ is described by:

$$I_s = |\sum_{i=1}^{N} C_i \, \varepsilon_0 \, \chi^{(2)}_{xxx} \cos(3\theta_i - 2\varphi)|^2 . \qquad (3)$$

We can therefore employ Eq. (3) with N= 2, using the total SHG from a pair of layers, to deduce the actual armchair orientation of the second SL by considering the first SL as a reference. This calculation can be performed provided that there is non-zero SHG signal and that the SLs are built as shown in Fig. 3, where part of the first SL at the bottom (SL1) is not overlapping with the second (SL2). In particular, the P-SHG signal detected from the uncovered region of SL1 can be used to calculate the armchair $\theta_1$ for SL1 (Eq. (3) with N= 1). Subsequently, the armchair orientation $\theta_2$ of the second SL2 (red area in Fig.3), can be derived upon using the known $\theta_1$ and Eq. (2). In this case, the resolution of $\theta_2$, calculated as error propagation of $\theta_1$ resolution 0.15°, is ~0.33°. In the same manner, by knowing $\theta_1$, $\theta_2$, the armchair of a third SL3 can be computed upon using again Eq. (3) for N= 3. This procedure can be repeated for an arbitrary number *N* of SLs, provided that there is always a non-overlapping region among the stacked layers (Fig. 3).

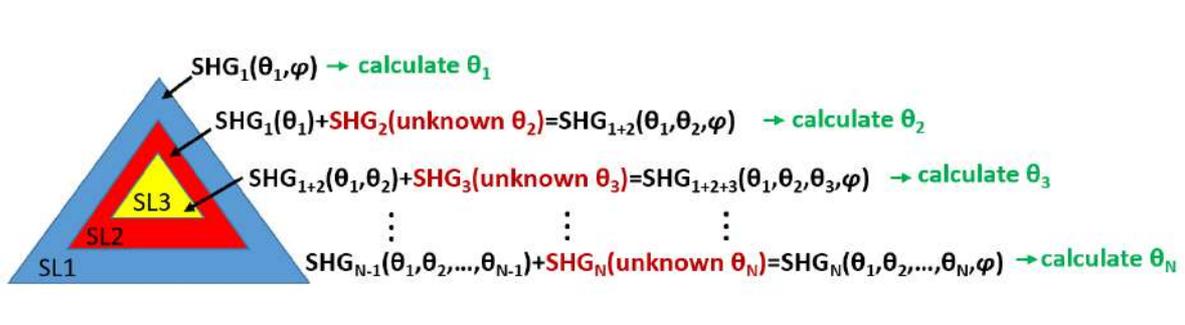

**Figure 3:** Schematic of three stacked SLs (SL1-3) partially overlapping. The procedure that enables the calculation of the armchair orientation of each individual SL is also presented. This process can be repeated for arbitrary number of SLs, as long as there is always a non-overlapping region among the stacked layers.

For example, in Fig. 4, we present the theoretical P-SHG modulation of two stacked SLs, as well as the modulation of their SHG interference signal in the overlapping region, for several twist angles of interest, assuming $A_1 = A_2$ (i.e., layers of equal SHG intensity). This assumption is realistic since layers of the same symmetry and similar composition should possess similar $\chi^{(2)}$. The blue polar diagrams shown in Fig. 4 are fixed and correspond to the SHG from an individual SL (Eq. (3) for N= 1, $A_1$= 1 and $\theta_1$= 0°), whereas the green polar diagrams show the SHG from a second SL (Eq. (3) for N= 1, $A_2$= 1, and varying armchair orientation $\theta_2$ from 0° to 60° with step 10°. Finally, the red polar diagrams in Fig. 4 correspond to the SHG interference from the two SLs in their overlapping area and were calculated by using Eq. (2) for $\theta_1$= 0° and $\theta_2 \in [0°, 60°]$ with step 10°.

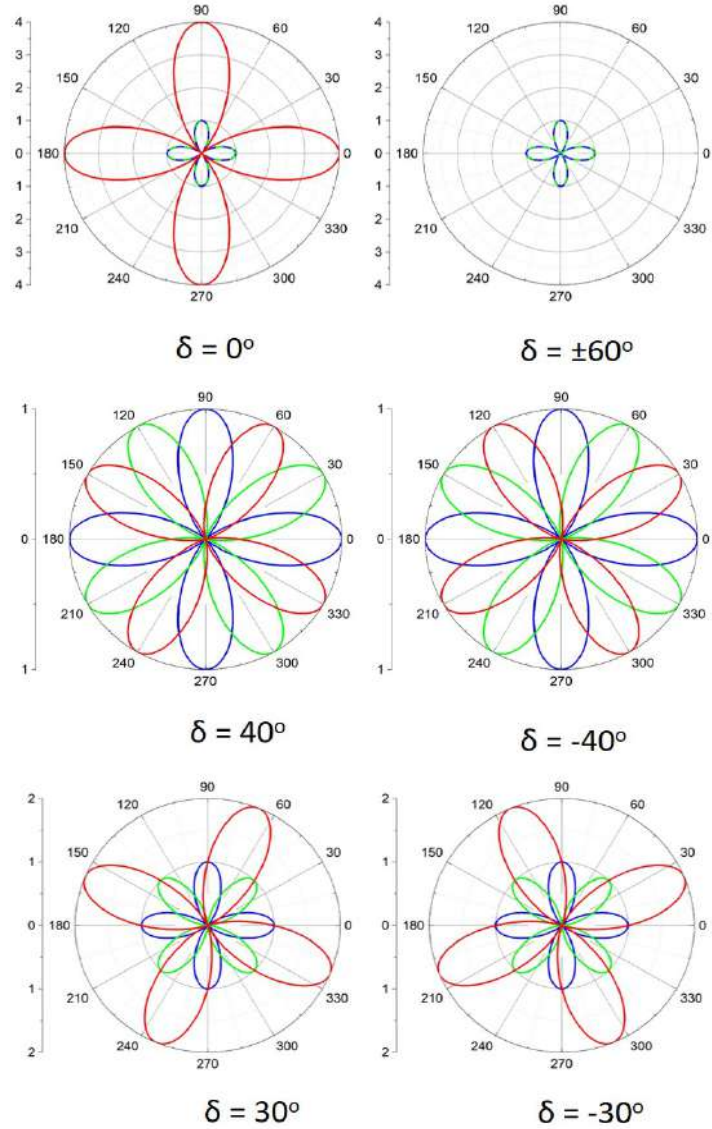

**Figure 4:** Simulated P-SHG from two stacked, SLs (blue and green polar diagrams), which act as surface SHG antennas radiating in phase. By changing the difference in the armchair orientation of the layers, control over the intensity and polarization of the SHG interference (red polar diagram) can be achieved. Note that the polar diagrams exhibit different SHG intensity scale.

Note that for $\delta = \pm 40°$, in Fig. 4, all three polar diagrams, corresponding to the SHG signal from SL1, SL2 and their overlapping area, reach the same maximum intensity. Upon rotating the excitation linear polarization $\mathbf{E}(\varphi)$, the SHG from SL1 (blue polar) reaches its maximum first, while the SHG signals from both SL2 (green polar) and the overlapping region (red polar) are zero. Subsequently, the SHG from both SL1 and SL2 goes to zero while that from the overlapping region maximizes. Finally, both the SHG from SL1 and the overlapping region go to zero as the SHG from SL2 becomes maximum. This means that in the case of two overlapping layers at $\delta = \pm 40°$, one could selectively: (i) switch on the SHG from SL1, while that of the overlapping region and SL2 are switched–off, or (ii) switch–on the SHG from the overlapping region, while that of SL1 and SL2 are both switched–off or (iii) switch–on the SHG from SL2, while that of the overlapping region and SL1 are both switched–off. Additionally, we note in Fig. 4 that when $\delta = \pm 30°$ the polar diagram produced by the interference of the two SLs (red line), passes from the maximum of the two polar diagrams produced by each individual layer SL1, SL2 (blue and green lines, respectively). Note also that when the SHG from the overlapping region is equal to the SHG from SL1 (SL2), the SHG from SL2 (SL1), is zero. This results in the complete switching–on of the SHG from one layer, while the SHG from the other layer is completely switched–off. This is also confirmed experimentally below, as for $\delta = \pm 30°$ one can selectively switch–on the SHG from the SL of preference.

Based on the above analysis, the amplitude of the P-SHG modulation (see Eq. (2)) originating from layered regions depends on both the number of SLs and their relative armchair orientation, i.e., the twist angle $\delta$. Consequently, a change in the SHG amplitude in a SL, is an indication of either the presence of a second TMD SL, or a change in the stacking order of the same SL (e.g. from 2H to 3R stacking[35]).

Figure 5a for $\varphi = 64°$ shows the SHG signal originating from one SL for the ROI shown in Fig. 1a, whereas Fig. 5b for $\varphi = 17°$ presents the SHG from an assumed second SL (see Eq. (2)). Finally, Fig. 5c is the summation of the two SHG images. In order to interpret the SHG signal variations and identify the existence of a second SL we perform a SHG-intensity profile analysis along the line-of-interest (LOI) in Figs. 5a-c. In particular, in Figs. 5d-f we plot the intensity profile for each pixel along this line. In Fig. 5a we mark with a white dashed line the region where the intensity drops close to zero. We set that lack of SHG signal as reference corresponding to the substrate. In Fig. 5b we mark with a light blue dashed line the region of the assumed second SL2. Finally, in Fig. 5c we note the overlapping region, contained within the yellow dashed line, as well as the empty and SL2 areas enclosed within the white and light blue dashed lines, respectively. We finally note the yellowish change in the colour in the overlapping region caused by the addition of green and red colours of Fig. 5(a) and Fig. 5(b), respectively. In Fig. 5d the SHG intensity everywhere along the LOI corresponds to that of one SL (~150 a.u.), except from the dark region (ranging from 200 to 350 pixels) where the material is absent and the SHG drops to zero. In Fig. 5e the SHG intensity again corresponds to one SL (~150 a.u. for pixels ~200 - ~500). The summation of the profile intensities shown in Figs. 5d-e, results in the profile intensity shown in Fig. 5f, where it is obvious that for pixels 0 - ~350 and ~500 - ~600 the SHG intensity corresponds again to one SL (~150 a.u.). Interestingly, the SHG intensity in the region where it is assumed that the two SLs overlap (pixels ~350 - ~500), is double (~300 a.u.) with respect to the one SL region, indicating the presence of the second SL, or a change in the stacking order. A change from 3R to 2H stacking is excluded, considering that it would give rise to a reduction of the SHG signal intensity[32]. It is therefore concluded that the increase of the SHG signal observed in the overlapping region, is a signature of an additional SL or a change to 3R stacking order (presence of second SL with equal armchair orientation) within the same layer.

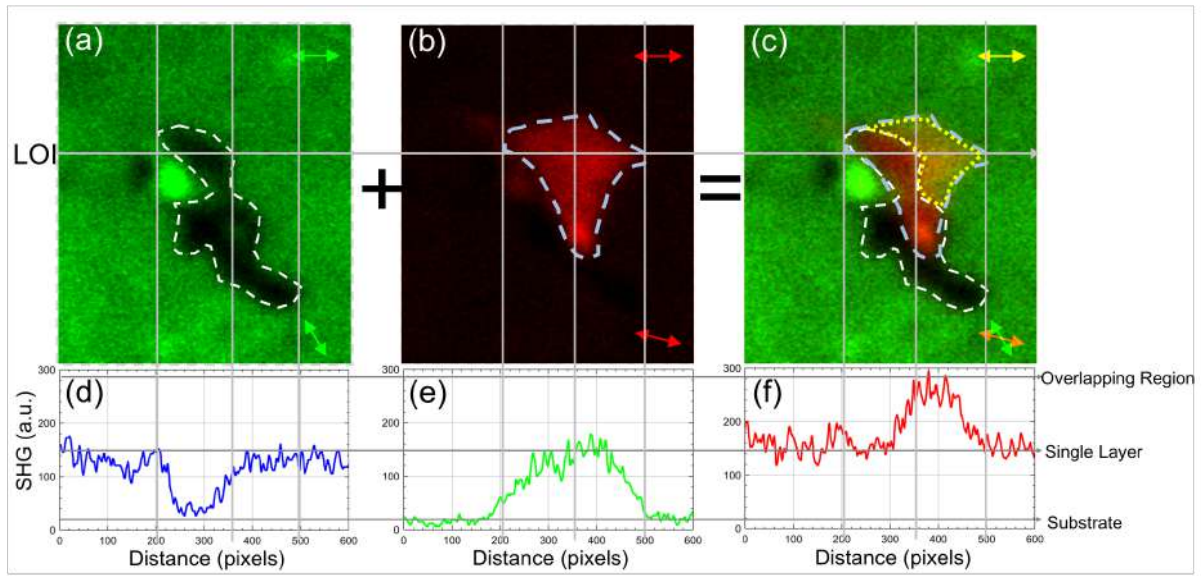

**Figure 5:** P-SHG imaging of the ROI in Fig.1a for analyzer parallel to the X-lab axis. Depending on the polarization angle of the excitation field the SHG originates from only (a) the lower ($\varphi$= 64°) or (b) the upper ($\varphi$= 17°) SL. In (a) the white dashed line encloses a region where material is lacking, whereas in (b) the light blue dashed line encircles the SL2 area. In (c) the two figures are superimposed, highlighting their overlapping area (encircled by the yellow dashed line). In (d)-(f) we plot the SHG intensity profile for the line of interest (LOI) in (a)-(c). In these plots, we identify the region of no SHG signal, assumed to be the substrate, two SL regions of equal intensity and the overlapping region where the SHG intensity is doubled.

In order to rule out the possibility of a change in the stacking order, intensity-only measurements are not sufficient and one has to perform a P-SHG analysis. Figs. 6a-d present the corresponding P-SHG polar plots for the same ROI as above and the armchair angles for points of interest (POI) lying in the different regions (SL1, SL2 and overlapping). Indeed, from the P-SHG signal obtained from POI1, represented with the blue polar diagram in Fig. 6b, we can calculate, using Eq. (3) with N= 1, the armchair angle of the bottom SL1 layer to be $\theta_1$= 39.5°. Using Eq. (2) with $\theta_1$= 39.5°, we obtain for POI2 of Fig. 6a the green polar shown in Fig. 6c, which corresponds to armchair angle of $\theta_2$= 12.8° for the SL2. Notably, the P-SHG signal from POI3 corresponds to the data points and the fitted red polar in Fig. 6d which leads, using Eq. (2) to $\theta_{eff}$= 25.1°. This signal is the result of the interference between the two SLs in POIs 1 and 2. By using the measured $\theta_{eff}$= 25.1° and by fixing to the measured $\theta_1$= 39.5°, we can calculate, using $\theta_{eff}=(\theta_1+\theta_2)/2$, that $\theta_2$= $2\theta_{eff}$ -$\theta_1$= 10.7°, which corresponds to the armchair angle of the second SL that produces the SHG interference (see Video S2). Consequently the twist angle can be calculated using two measurements, one in the overlapping and another in the monolayer region, as $\delta$= 2($\theta_1$- $\theta_{eff}$) =28.8°.

From the above observations we can exclude the possibility of a change in the stacking order within the same SL in the overlapping region, since in that case $\theta_{eff}$ should be equal to $\theta_1$ (3R stacking originates in lattices of the same armchair orientation). Indeed, using P-SHG measurements we calculate a $\theta_{eff}$ that indicates the presence of a second SL with armchair orientation that fits very well with that calculated for the overhanging SL2 region. These results denote a continuous region of the same armchair orientation and thus the presence of a second SL. This region can also be optically isolated in the SHG image of Fig. 5b. It should be noted here that the four-leave rose patterns shown in Figs. 6b-d correspond to the P-SHG signatures of $WS_2$[9]. Therefore, any detected SHG signal that does not comply with this pattern modulation, is excluded during the data fitting.

It should also be emphasized that the experimental twist angle calculated above $\delta$=28.8° in the overlapping region, is very close to the theoretical one $\delta$= 30° required for complete suppression of the SHG signal from one SL. Confirming our theoretical prediction (see Fig. 4), the experimentally inferred $\delta$=28.8°, results in the almost complete suppression of the SHG from one layer, while the SHG of the other is maximum, as it is experimentally demonstrated in Fig. 1 and Video S1. This is also shown in Fig. 6d and Video S2, where the experimental polar diagram produced by the interference of the two SLs (red curve) passes very close to the maximum of the two polar diagrams produced by each individual layer (blue and green curves). Note that when the SHG interference intensity is equal to the maximum of the SHG intensity from one SL, the SHG intensity from the other SL is almost zero. This results in the suppression of the SHG from the first layer, while at the same time the SHG from the second layer is completely switched–on and vice-versa. It is stressed out that the second SL was not intentionally placed in the ideal 30° twist angle but it was naturally grown during the CVD, measured at $\delta=28.8°$ twist angle.

In this context, we can interpret in detail the P-SHG map obtained from the multilayered CVD-grown $WS_2$ triangular flake shown in Fig. 1 as follows. Fig. 6e presents the map of armchair angles θ for the ROI-1 seen in Fig. 1a. This map is obtained by performing pixel-wise fitting using Eq. (3) for N= 1 (assuming only one layer). Using this map one can create the armchair histogram of specific areas, for example the one shown in Fig. 6g. Notably, the armchair mapping gives values of θ ~ 40° for the outer region, but only θ ~ 25° for the central one (yellowish). In addition in Fig. 6e, there is a part of the central region that exhibits θ ~ 12° (blue). The above results confirm the presence of more than one SL. If we now assume two different SLs and fix the measured $\theta_1$= 39.5° in the monolayer and use the measured $\theta_{eff}$ in the overlapping region, we obtain the map of the twist angle shown in Fig. 6f and its corresponding armchair histogram shown in Fig. 6h. In this new histogram $<\delta>$= 30.12° and σ= 3.44°.

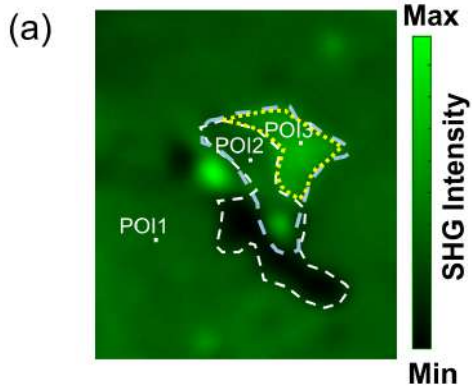
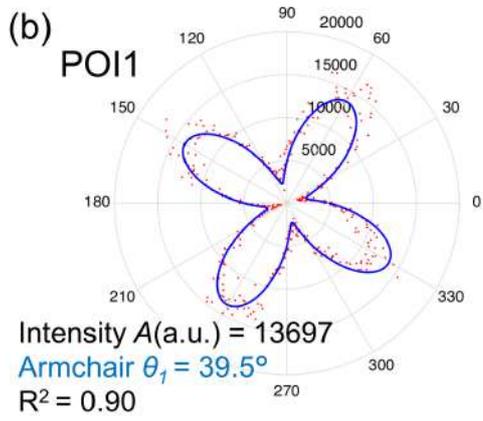
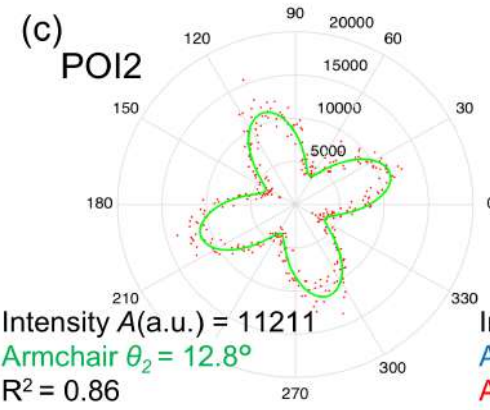
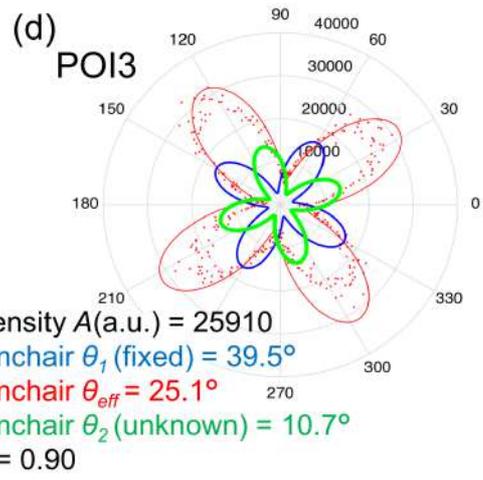
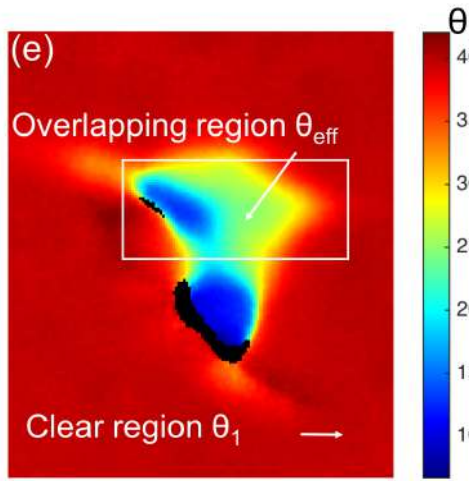
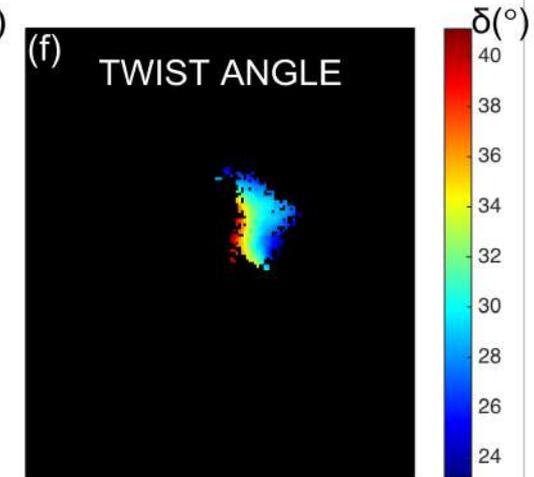
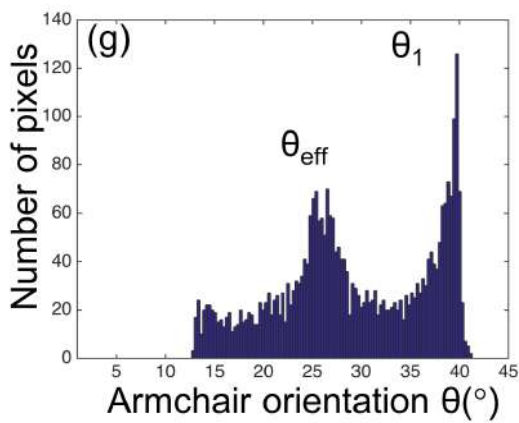
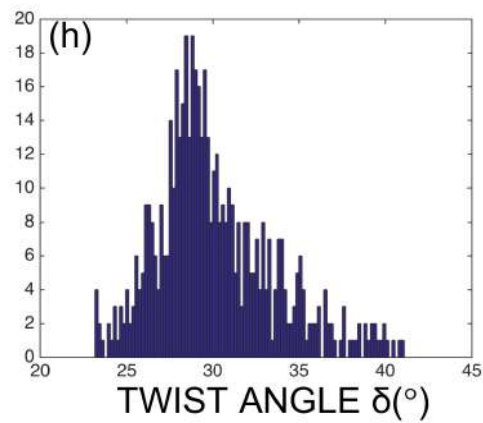
13

**Figure 6: Twist angle mapping in the overlapping region of two SLs in CVD-grown WS$_2$** (a) Total SHG without analyzer where three POIs are indicated. (b) PSHG polar diagram with respect to the excitation linear polarization $\varphi \in [1°, 360°]$, step 1° for POI1. By fitting the data with Eq. (3) for N= 1 we obtain the blue polar diagram which corresponds to armchair $\theta_1$= 39.5°; (c) For POI2 fitting again the data with Eq. (3) for N= 1 we obtain the green polar diagram that corresponds to armchair $\theta_2$= 12.8°; (d) PSHG from POI3 (red dots) and fitted polar diagram (red curve) corresponding to armchair $\theta_{eff}$= 25.1°. This is the result of the interference between the two, SLs shown in the POIs of (b) and (c). By using the measured $\theta_{eff}$= 25.1° and by fixing $\theta_1$= 39.5°, we calculate $\theta_2$= 10.7° (using $\theta_2$= 2$\theta_{eff}$ - $\theta_1$), which corresponds to the armchair of a second SL that produces the SHG interference (see Video S2). (e)-(h) Pixel-wise mapping of crystal orientation: (e) Fitting with Eq. (3) for N= 1; (f) Twist angle mapping using $\delta$= 2($\theta_1$- $\theta_{eff}$); (g), (h) Image histograms of the armchair values showing the experimentally retrieved values $\theta_1 \sim$ 39.5° and $\theta_{eff} \sim$ 25.1° originating from the SHG interference between the two overlapping SLs, as well as twist angle mapping in the overlapping region. The twist angle distribution presented <$\delta$>= 30.12° with $\sigma$= 3.44°.

We applied the above analysis in the simplest case of an artificially prepared WS$_2$ homobilayer (see Methods and Video S3), presented in Fig. 7. Our technique readily mapped the twist angle in the overlapping region of the two monolayers (see Fig. 7h). Fig. 7a shows the total SHG intensity from a large area of two WS$_2$ monolayers overlapping in a region. Focusing on a smaller ROI, that contains both overlapping and monolayer regions, we obtain in Fig. 7b the corresponding SHG within the region defined by the white rectangle. Choosing three points of interest (POIs) within the ROI we present in Figs. 7c-f the polar diagrams of the SHG intensity as function of the polarization angle of the fundamental field. Repeating the same procedure for all the pixels inside the ROI we obtain in Fig. 7g the armchair angle mapping and in Fig. 7h the corresponding armchair angle distribution within the ROI. Finally using the procedure described above we deduce the twist angle mapping and its corresponding distribution, shown in Figures 7i and 7j, respectively.

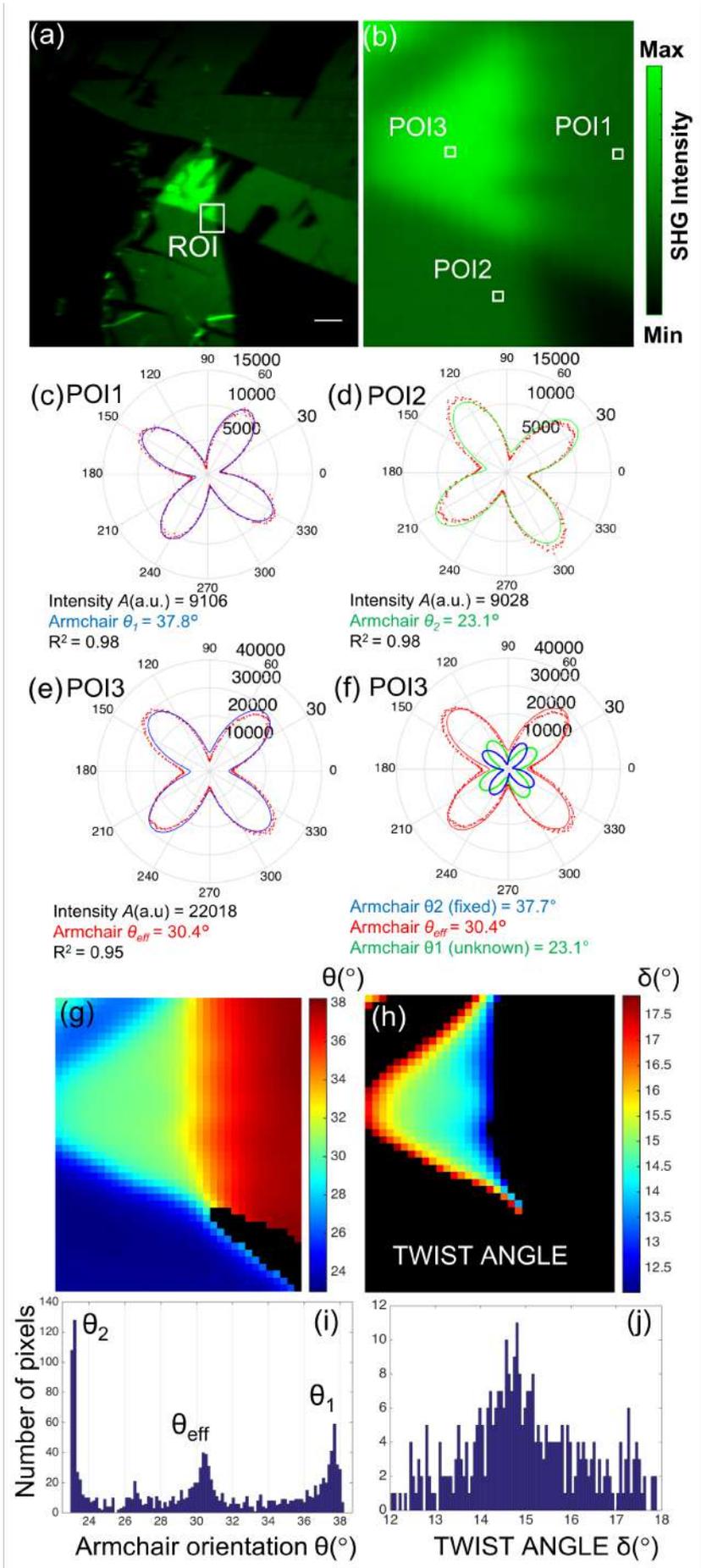

**Figure 7: Twist angle mapping in the overlapping region of WS$_2$ twisted bilayer.** (a) Total SHG (without analyzer) of two WS$_2$ monolayers overlapping in a region of interest (ROI). Note the increase of the SHG in the overlapping region. Scale bar is 10 μm. (b) Enlarged image of the total SHG intensity within the ROI selected in (a). (c)-(f) Polar diagrams of SHG intensity from three POIs indicated in (b). (g) Mapping of armchair orientation in ROI. (h) Twist angle spatial mapping of the overlapping region of the two WS$_2$ monolayers. (i) Image histogram of (g) showing the armchair angles distribution. (j) Image histogram of (h) showing the twist angles distribution with $<\delta>$=14.9° and σ=1.3°.

In particular, the measured armchair angle of SL1 in the non-overlapping region is $\theta_1$ (measured, POI1 in Fig. 7b) = 37.8° (Fig. 7c), whereas the effective armchair angle in the overlapping region is $\theta_{eff}$ (measured, POI3 in Fig. 7b) = 30.4° (Fig. 7e). The overlapping region in the region of interest (ROI) seen in Fig. 7a, provided the twist angle mapping seen in Fig 7h. This results in $\theta_2 = 2\theta_{eff} - \theta_1 = 23.0°$ (Fig. 7f). This value is indeed very close to the experimentally measured one in the non-overlapping region $\theta_2$ (measured, POI2 in Fig. 7b) = 23.1° (Fig. 7d). We can finally extract the mean of the distribution of the twist angle values seen in Fig. 7h, using $<\delta> = 2(\theta_1 - \theta_{eff}) = 14.9°$, σ = 1.3° (Fig. 7j).

**Conclusions**

In conclusion, we have used P-SHG imaging microscopy to map the twist angle in stacked TMD layers. In particular, the effect of the SHG interference in the overlapping region of two individual WS$_2$ monolayers was described in terms of the newly introduced concept of the effective orientation $\theta_{eff}=(\theta_1+\theta_2)/2$ which dictates the P-SHG modulation in an overlapping region of two stacked twisted 2D layers. This novel concept of the effective orientation allowed the determination of the armchair orientation of a second layer that contributed to the total SHG signal detected in the overlapping region. Consequently, by determining the crystal orientation of the second layer that resulted to the measured SHG in the overlapping region we were able to calculate the twist angle between the two layers and for the first time create its pixel-by-pixel mapping both in CVD-grown and in artificially stacked WS$_2$ bilayers. Thus, we have demonstrated an all-optical technique that identifies the presence of stacked SLs, calculates and maps their twist angle. In addition, we have shown experimentally and interpreted theoretically, that when the twist angle between the two SLs is 30° one can selectively suppress the SHG from one layer, while at the same time the SHG from the other is switched-on and vice-versa. This SHG "magic" twist angle enables axial super-resolution imaging and consequently provides a quality characterization of the layered 2D-structure. We envisage that our methodology will provide a new and easy characterization tool of twisted 2D crystals towards their numerous optoelectronic applications.

## Methods

**Custom-Built polarization-in, polarization-out SHG microscope.** Our experimental apparatus is based on a diode-pumped Yb:KGW fs oscillator (1030 nm, 70–90 fs, 76 MHz, Pharos-SP, Light Conversion, Vilnius, Lithuania) inserted in a modified Axio Observer Z1 (Carl Zeiss, Jena, Germany) inverted microscope (Fig. 2). The laser beam is passing through a zero-order half-wave retardation plate (QWPO-1030-10-2, CVI Laser), placed at a motorized rotation stage (M-060.DG, Physik Instrumente, Karlsruhe, Germany) that rotates with high accuracy (1°) the orientation of the excitation linear polarization. Raster-scanning of the beam at the sample plane is performed using a pair of silver-coated galvanometric mirrors (6215H, Cambridge Technology, Bedford, MA, USA). The beam is reflected on a silver-coated mirror, at 45° (PFR10-P01, ThorLabs, Newton, NJ, USA), placed at the motorized turret box of the microscope, just below the objective (Plan-Apochromat 40x/1.3NA, Carl Zeiss). The choice, of the silver coating of all the mirrors (PF 10-03-P01, ThorLabs), including the galvanometric mirrors, makes our setup insensitive to the laser beam polarization and its angle of incidence. The mean polarization extinction ratio of the different linear polarization orientations, calculated using crossed polarization measurements at the sample plane, was 28:1. In the forward direction, the SHG is collected using a high numerical aperture (NA) condenser lens (achromatic-aplanatic, 1.4NA, Carl Zeiss). The SHG is separated from the laser using a short-pass filter (FF01-720/SP, Semrock, Rochester, NY, USA) and from any unwanted signal using a bandpass filter (FF01-514/3, Semrock).  A rotating film polarizer (LPVIS100- MP, ThorLabs) is placed just in front of the PMT (H9305-04, Hamamatsu, Hamamatsu City, Japan) to measure the anisotropy due to the polarization of the SHG signals. The P-SHG imaging (i.e. in the exfoliated $WS_2/WS_2$ stacked structure) is performed in the epi-detection using a dichroic mirror (DMSP805R, ThorLabs) and the forward P-SHG detection module described above, placed in an epi-detection port of the microscope. Coordination of PMT recordings with the galvanometric mirrors for the image formation, as well as the movements of all the motors, is carried out using LabView (National Instruments, Austin TX, USA) software.

**Samples.** The $WS_2$ samples were grown by the low-pressure chemical vapor deposition method (LP-CVD) on a *c*-cut (0001) sapphire substrate (2D Semiconductors). Note that in the case of CVD-grown samples the stacking of layers is not artificial like in [12] but occurs naturally during the growth[35]. Nevertheless, we expect a similar behavior like in [35] from CVD-grown layered samples. This effect has been observed previously and is attributed to the nucleation that occurs during the CVD-growth and commences from the center[36].

WS2 bulk crystals were exfoliated by micromechanical cleavage on a polydimethylsiloxane (PDMS) stamp, placed on top of a glass slide for optical inspection. The first monolayer was transferred on a $Si/SiO_2$ (285nm) substrate by an all-dry viscoelastic stamping and then it was mounted on a XYZ micromechanical stage[37]. The stage was placed under a coaxially illuminated microscope and following the same procedure, a second $WS_2$ monolayer on a different PDMS was carefully aligned and then stamped slowly on top of the first monolayer. The final step included a controlled release of the PDMS stamp for the fabrication of the $WS_2$ bilayers.

**Data availability**

The data that support this study are available from the corresponding author upon reasonable request.

**Conflict of Interest**

The authors declare that there are no competing interests.

**Acknowledgements**

This work is supported by the European Research Infrastructure NFFA-Europe, funded by the EU's H2020 framework program for research and innovation under grant agreement n. 654360. S. P. and I. P. acknowledge financial support from the Stavros Niarchos Foundation within the framework of the project ARCHERS ("Advancing Young Researchers' Human Capital in Cutting Edge Technologies in the Preservation of Cultural Heritage and the Tackling of Societal Challenges").

**Author contributions**

SP, ES and GK planned the project; SP, LM and IP designed the experiment; SP and IP conducted the optical experiments; SP, AL and LM conducted the data analysis; AL provided technical support; GKou prepared the exfoliated samples. LM and SP elaborated the theoretical model; ES and GK guided the research. All authors contributed to the discussion and preparation of the manuscript.

**Supplementary Information**

Video S1, Video S2, Video S3.